\documentclass[%
 reprint,
superscriptaddress,
floatfix,
 amsmath,amssymb,
 aps,
 prx,
longbibliography
]{revtex4-2}

\usepackage{graphicx}
\usepackage{dcolumn}
\usepackage{bm}
\usepackage{placeins}
\usepackage{float}
\usepackage{hyperref}
\usepackage{placeins}
\usepackage{times}
\usepackage{physics}
\usepackage{xcolor}
\usepackage{float}
\usepackage{hyperref}
\hypersetup{
    colorlinks,
    citecolor=blue,
    filecolor=black,
    linkcolor=black,
    urlcolor=black
}
\usepackage[title,toc]{appendix}

\begin{document} 

\title{Mid-circuit correction of correlated phase errors using an array of spectator qubits}


\author
{\normalsize{K. Singh,$^{1,\dag}$ 
C. E. Bradley,$^{2,\dag}$
S. Anand,$^{2,\dag}$ 
V. Ramesh,$^{3}$ 
R. White,$^{3}$ 
and H. Bernien$^{2\ast}$}\\
\small{$^{1}$Intelligence Community Postdoctoral Research Fellowship Program,}\\
\small{Pritzker School of Molecular Engineering, University of Chicago, Chicago, IL 60637, USA}\\
\small{$^{2}$Pritzker School of Molecular Engineering, University of Chicago,  Chicago, IL 60637, USA}\\
\small{$^{3}$Department of Physics, University of Chicago, Chicago, IL 60637, USA}\\
\small{$^\dag$These authors contributed equally to this work.} \\
\small{$^\ast$To whom correspondence should be addressed; E-mail:~bernien@uchicago.edu.}
}

\date{\today}


\begin{abstract}

Scaling up invariably error-prone quantum processors is a formidable challenge. While quantum error correction ultimately promises fault-tolerant operation, the required qubit overhead and error thresholds are daunting, and many codes break down under correlated noise. Recent proposals have suggested a complementary approach based on co-located, auxiliary ‘spectator’ qubits. These act as in-situ probes of noise, and enable real-time, coherent corrections of the resulting errors on the data qubits. Here, we use an array of cesium spectator qubits to correct correlated phase errors on an array of rubidium data qubits. Crucially, by combining in-sequence readouts, data processing, and feed-forward operations, these correlated errors are suppressed within the execution of the quantum circuit. The protocol is broadly applicable to quantum information platforms, and our approach establishes key tools for scaling neutral-atom quantum processors: mid-circuit readout of atom arrays, real-time processing and feed-forward, and coherent mid-circuit reloading of atomic qubits.

\end{abstract}

\maketitle

\section*{Introduction}

Realizing large-scale programmable quantum systems that can overcome inevitable noise sources is a central challenge for modern science \cite{Campbell2017,Terhal2015}. Environmental noise and experimental parameter drift necessitate strategies to reduce their impact and overcome resulting qubit errors. While quantum error correction will ultimately be required, achieving the necessary qubit operation fidelities is an outstanding challenge for present quantum computing platforms~\cite{Ryananderson2021,Postler22,Zhao2022,Krinner2022,Acharya2022,Abobeih2022,Bluvstein22}. Moreover, the effectiveness of error correction codes is reduced by correlated errors \cite{Preskill2012,Fowler14}, which may naturally occur when the qubits are in close spatial proximity or are controlled by shared hardware \cite{Kuhr05, Monz14, Bradley19, Boter20, Wilen21}.

To address these challenges, a number of techniques have been developed to mitigate the effects of noise, such as composite pulses \cite{Vandersypen2005}, optimal control \cite{Rabitz2004}, dynamical decoupling \cite{Gullion1990,Vandersypen2005}, Hamiltonian learning \cite{Shulman2014}, and machine-learning-based control engineering \cite{Mavadia2017}. While these techniques have found great success, they are typically tailored to specific noise models or require careful calibration, and thus face challenges when employed in realistic, fluctuating environments. For example, dynamical decoupling generates a filter function which mitigates a particular spectrum of noise, with pass-bands remaining that are not suppressed~\cite{Degen17}. Additionally, it is only effective if the correlation time of the noise is long with respect to the interpulse delay. 

Recent theoretical work has proposed a complementary technique based on `spectator' qubits: additional qubits which are co-located with the computational `data' qubits and are susceptible to the same noise sources. Spectator qubits act as in-situ probes of that noise, such that measurement and feed-forward can be used to coherently protect the data qubits during the execution of a quantum algorithm~\cite{Majumder2020,Gupta20b,Song2022}. Notably, under two key conditions, spectator protocols are agnostic to the spectrum and correlation time of the noise source. First, the noise-induced dynamics must be correlated between the spectator and data qubits. Second, an estimate of those dynamics must be made by reading out the spectator qubits --- and a subsequent feed-forward operation applied --- much faster than the timescale over which the data and spectator qubits decorrelate. This second requirement has limited the experimental implementation of such protocols, as a significant number of measurements are required to reliably estimate the effects of a dynamic noise environment. Furthermore, the spectator qubit readouts must be performed mid-circuit without perturbing the data qubits.

\begin{figure*}[ht!]
\centering
\includegraphics{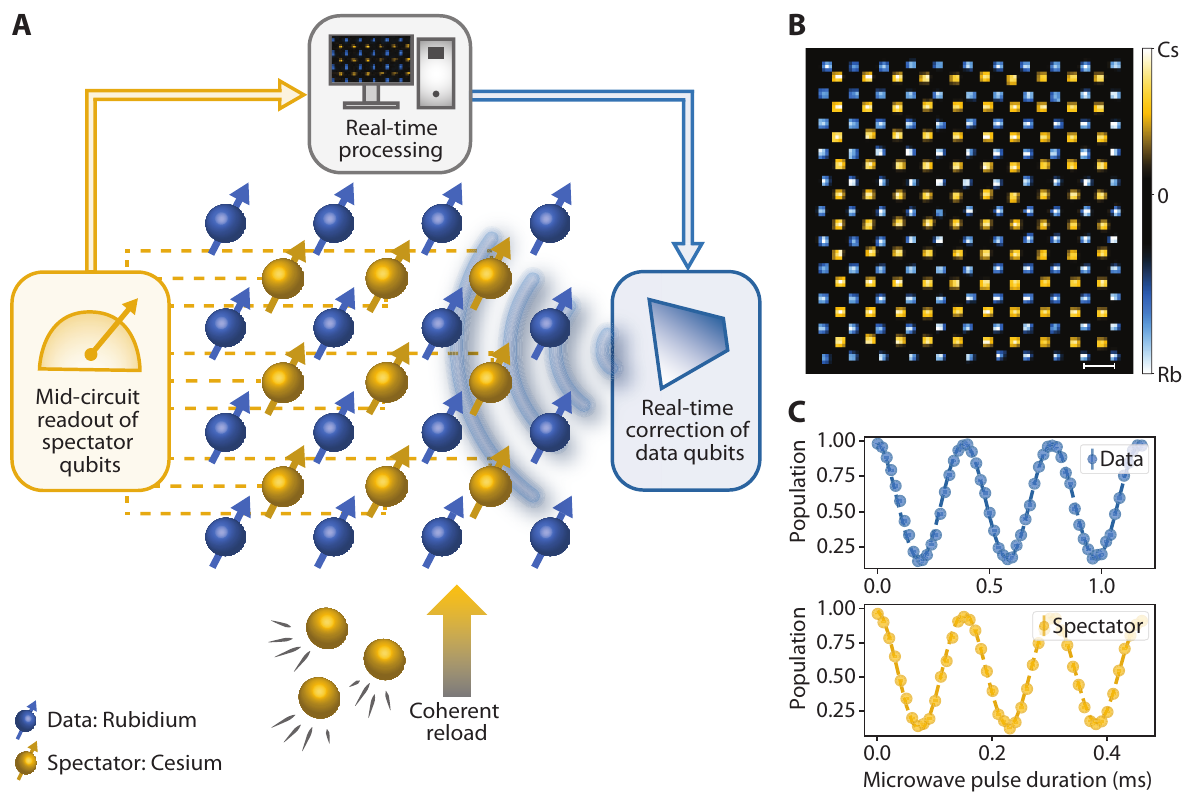}
\caption{\textbf{Spectator qubit protocol with a dual-species atom array.} (\textbf{A}) Feed-forward loop for real-time correction of correlated phase errors between data qubits (Rb atoms, blue) and spectator qubits (Cs atoms, yellow). A mid-circuit, single-shot phase estimation on the spectators is used to infer the noise-induced phase accrued by the data qubits. This information enables a real-time correction on the data qubits prior to the final readout, suppressing dephasing. Subsequently, spectator qubits lost during readout can be replenished while maintaining data qubit coherence. (\textbf{B}) Example fluorescence image of the dual-species atom array. Scale-bar indicates $\sim10~\mu m$. (\textbf{C}) Microwave Rabi oscillations of the data and spectator qubits. Dashed lines are fits to exponentially decaying sinusoids.}
\label{fig:Schematic}
\end{figure*}

Here, we overcome these challenges and demonstrate real-time correction of correlated phase errors using a dual-species array of individually trapped neutral atoms. The protocol is outlined in Fig.~\ref{fig:Schematic}A.  Data qubits (rubidium atoms) and spectator qubits (cesium atoms) are laser-cooled into optical tweezer arrays~\cite{Singh2022}. During logic operations on the data qubits, mid-circuit readouts on the array of $\sim$ 60 spectator qubits enable single-shot estimation of globally correlated phase errors. The readout results are processed in real-time and used to infer the noise-induced phase accrued by the $\sim$ 60 data qubits. Crucially, due to the crosstalk-free operation of the two species, these readouts do not disturb the coherence of the data qubits. We leverage a classical control architecture to perform in-sequence feed-forward, such that correlated errors on the data qubits are mitigated within the execution of the quantum circuit. Finally, we show that the spectator qubits can be replenished within the data qubit coherence time, an essential step towards repeated measurements and the continuous operation of atom-based quantum processors.

Our experiment is performed on arrays of 10x10 and 11x11 sites for the spectator and data qubits respectively (Fig.~\ref{fig:Schematic}B), which are stochastically loaded with an average loading fraction of $\sim$ 55\%. The qubits are encoded into long-lived hyperfine states ($\ket{F = 1, m_F = 0} := \ket{0}_\text{Rb}$ and $\ket{F = 2, m_F = 0} := \ket{1}_\text{Rb}$ for Rb; $\ket{F = 3, m_F = 0} := \ket{0}_\text{Cs}$ and $\ket{F = 4, m_F = 0} := \ket{1}_\text{Cs}$ for Cs). Microwave driving of the data and spectator qubits after optical pumping into $\ket{1}_\text{Rb}$ and $\ket{1}_\text{Cs}$ reveals coherent Rabi oscillations (Fig.~\ref{fig:Schematic}C). 

\vspace{-5pt}
\section*{Results}
\subsection*{Mid-circuit readout of atomic qubits}

An essential ingredient for the spectator protocol is to perform mid-circuit readout (MCR) of the spectator qubits without inducing additional data qubit decoherence. This is challenging in single-species atom arrays, since all atoms are resonant with the excitation laser, and the measured qubits scatter light which can decohere the data qubits via reabsorption. To overcome this, several ideas have been proposed and demonstrated, including coherently transporting qubits into readout cavities~\cite{Deist22}, or using additional shelving states to hide atoms from excitations from the readout light, as demonstrated for trapped ions~\cite{Postler22}. However, realizing crosstalk-free imaging in large atom arrays has remained an outstanding challenge. A key motivation behind the dual-species approach is that the different atomic species have distinct optical transitions, and measurements on one species are not expected to influence the other~\cite{Singh2022, Zhang2022}.

\begin{figure*}[ht!]
\centering
\includegraphics{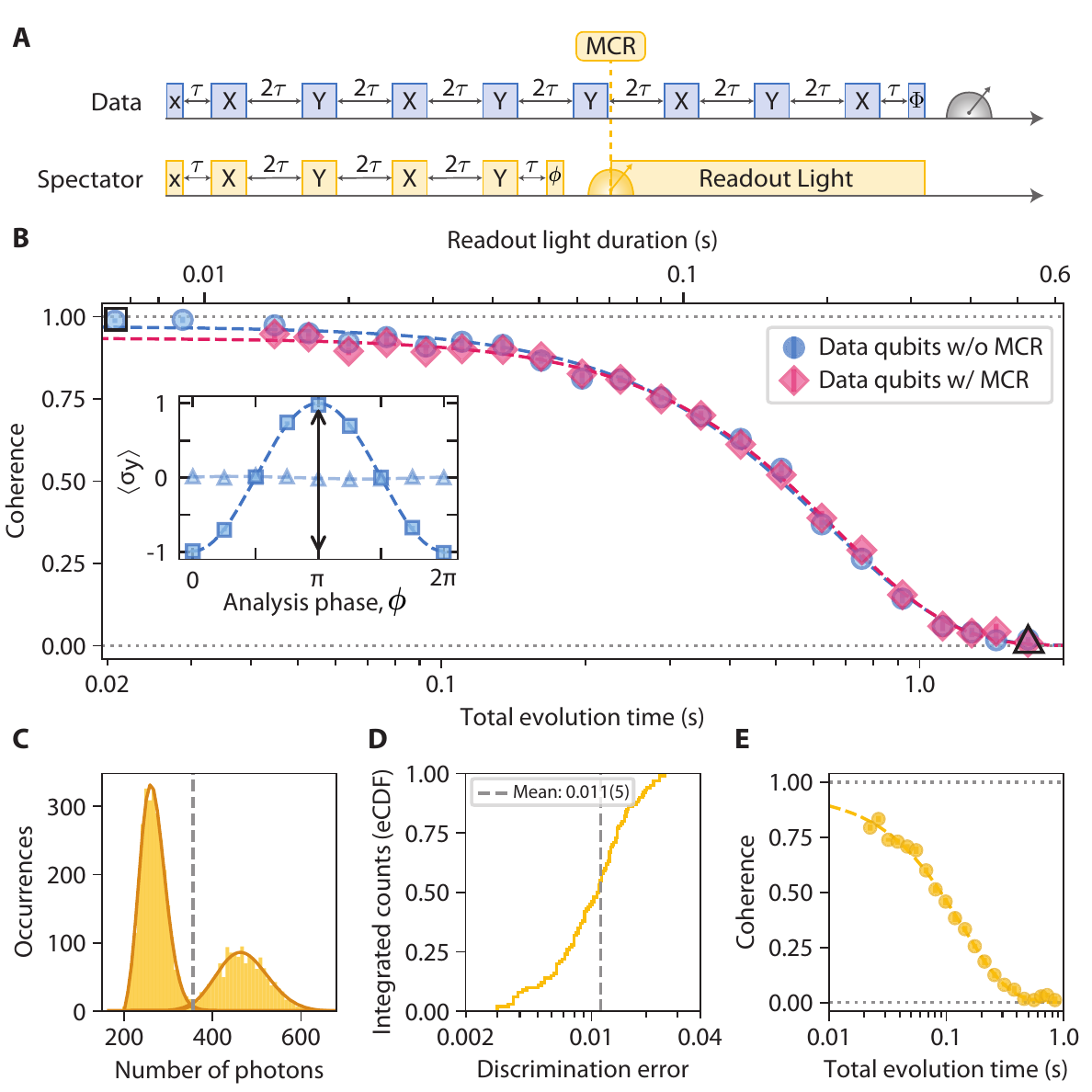}
\caption{\textbf{Mid-circuit readout of atomic qubits.} (\textbf{A}) Pulse diagram depicting mid-circuit readout (MCR) of atomic qubits. Lower case and upper case letters indicate $\pi/2$ and $\pi$ pulses, respectively, along that axis of rotation.  Readout light is left on for the remaining duration of the sequence after MCR. (\textbf{B}) Measurement of spectator qubit dynamics while preserving data qubit coherence. During an XY8 decoupling sequence on the data qubits (red diamonds), we perform an XY4 decoupling sequence and subsequent projective measurement on the spectator qubits. The data qubit coherence ($\sqrt{\langle\sigma_{x}\rangle^{2}+\langle\sigma_{y}\rangle^{2}}$) is unchanged in the absence of MCR (blue circles). Dashed lines are fits (see supplement). Inset: coherence measurements for early (square) and late (triangle) evolution times. (\textbf{C}) Example fluorescence histogram of a spectator qubit. Solid lines are fits to a bimodal Poisson distribution. (\textbf{D}) Cumulative histogram of the discrimination infidelities of the spectator qubits during mid-circuit readout (see supplement). (\textbf{E}) Coherence of spectator qubits. The measured spectator coherence time is $T_{2}^{\mathrm{XY4}} = 136(7)$~ms.}
\label{fig:MCR}
\end{figure*}

In a first experiment, we characterize the spectator qubit mid-circuit readout, and measure its impact on the data qubit coherence. The quantum circuit is shown in Fig.~\ref{fig:MCR}A. During an XY8 decoupling sequence on the data qubits, an XY4 sequence is performed on the spectators. The spectator qubits are measured within the XY8 sequence by selectively removing all atoms in the $\ket{1}_\text{Cs}$ state via a resonant laser pulse and then fluorescence imaging for 15 ms. The coherence of the data and spectator qubits as a function of their individual decoupling times are shown in Figs.~\ref{fig:MCR}B,E, respectively. While the camera exposure time is fixed, the imaging light is applied for a variable time, $5\tau$ (of a total of $16\tau$) in order to determine its effect on the data qubits. Crucially, the data qubit coherence time is unaltered by the MCR (fitted $T_{\mathrm{2, MCR}}^{\mathrm{XY8}}$ = 0.68(1)~s, $T_{\mathrm{2, No\ MCR}}^{\mathrm{XY8}}$ = 0.65(2) s). The large detuning of the imaging light leads to negligibly low spontaneous scattering rates of $\sim 10^{-7}$~Hz. Moreover, spontaneous Raman scattering events which change these $m_{F}$ states are further suppressed by a factor of 0.009 due to destructive interference of the off-resonant transition amplitudes \cite{Kuhr05}. The theoretical $T_{1}$ time from this decay channel is thus $\sim 10^{8}$~s, resulting in a data qubit bit flip rate from readout crosstalk of $\sim 10^{-11}$ during the 15~ms MCR. The discrimination fidelity of the spectator qubit states (Fig. ~\ref{fig:MCR}D) is extracted from a bimodal fit to the fluorescence histogram of each spectator qubit, as exemplified in Fig. ~\ref{fig:MCR}C (see supplement). Across the spectator array we find a mean fidelity of 0.989(5), showing that the spectator qubit states are well-resolved by MCR.

\vspace{-5pt}
\subsection*{Mid-circuit correction of correlated phase errors}

\begin{figure*}[ht!]
\centering
\includegraphics[width = \textwidth]{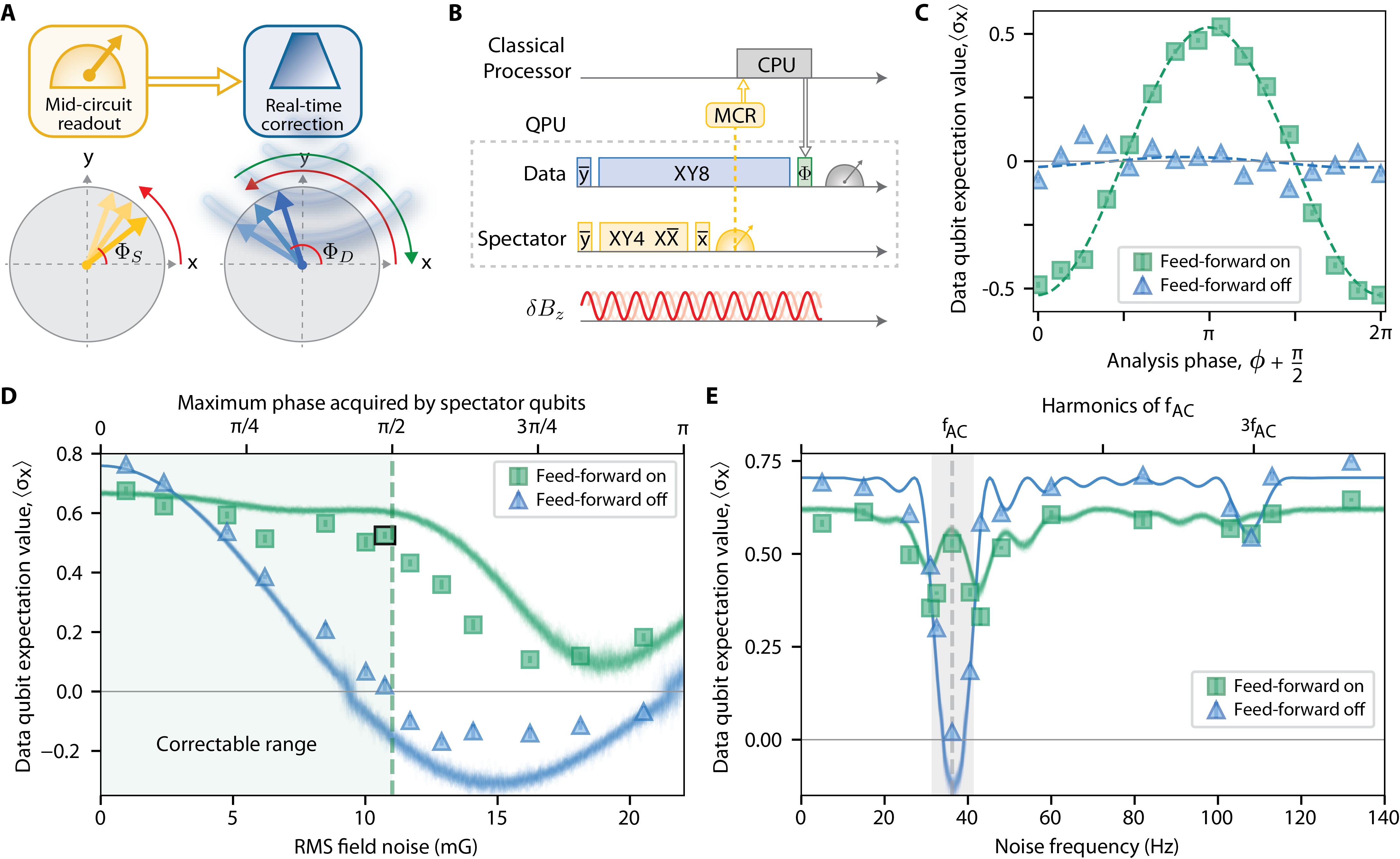}
\caption{\textbf{Mid-circuit correction of correlated phase errors.} (\textbf{A}) Noise channels induce correlated phase errors (red arrows) between the two sets of qubits. Measurement of the spectators along the y-axis enables single-shot phase estimation, from which the phase accrued by the data qubits can be inferred and corrected in real-time. (\textbf{B}) Gate sequence. The data and spectator qubits are synchronously decoupled and acquire correlated errors due to magnetic field noise $\delta B_z$. The spectator qubit decoupling sequence is truncated, with the remaining time assigned for mid-circuit readout and feed-forward. (\textbf{C}) Example coherence measurement of the data qubits at the end of the sequence, with the feed-forward turned on (green squares) and off (blue triangles). Field noise is applied at $f_{\mathrm{AC}} = 36.2$ Hz, 10.7 mG RMS. Dashed lines are fits, from which we extract $\langle \sigma_{x} \rangle$ = 0.53(1) and 0.02(2), respectively.  (\textbf{D}) Data qubit $\langle \sigma_{x} \rangle$ as a function of the RMS noise strength at $f_{\mathrm{AC}}$. The shaded green region indicates the correctable range (see text). (\textbf{E}) Data qubit $\langle \sigma_{x} \rangle$ as a function of the noise frequency at 10.7 mG RMS. Shaded grey region indicates an absolute gain in the measured coherence. For panels D and E, solid lines are the results of numerical simulations (see text).}
\label{fig:Feedback}
\end{figure*}

The preservation of data qubit coherence during spectator readout opens the possibility for feed-forward operations within a quantum circuit. Under simultaneous evolution, noise channels can induce correlated phase errors between the data and spectator qubits. Importantly, the large number of spectator qubits allows single-shot estimation of the acquired phase from one simultaneous MCR. The phase accrued by the data qubits can then be inferred and corrected in real-time, as illustrated in Fig.~\ref{fig:Feedback}A. 

To demonstrate this capability, we inject global magnetic field noise with amplitudes and frequencies comparable to those typically found in laboratory environments. The phase of the noise is random in each experimental repetition, without shot-to-shot temporal correlations.
We focus on monochromatic noise for ease of synthesis and interpretation of protocol performance, but note that our scheme is generally agnostic to the noise spectrum. The pulse sequence for the experiment is shown in Fig.~\ref{fig:Feedback}B. The data and spectator qubits undergo synchronous dynamical decoupling and acquire correlated errors from the common noise. While the filter function of the CPMG-type dynamical decoupling sequence partially mitigates such noise, certain frequencies still couple into the sequence, occurring at odd-harmonics of $f_{\mathrm{AC}} = 1/(4\tau) = 36.2$ Hz, where $2\tau$ is the time between $\pi$-pulses \cite{Degen17}. The spectators sample this noise for three-quarters of the total evolution time of the data qubits, with the remainder of the time assigned for MCR and feed-forward. To achieve fast camera processing and feedback, we utilize a camera-linked classical control architecture for in-sequence processing of the fluorescence images, which in turn triggers an arbitrary-waveform-generator to perform real-time updates of the phase of the final data qubit $\pi/2$-pulse (see supplement). The phase update of this final $\pi/2$-pulse is equivalent to a $Z$-axis qubit rotation, which is used to correct the noise-induced phase error on the data qubits. 

To estimate the phase acquired by the spectators, $\Phi_{S}$, MCR is performed along an axis orthogonal to the state preparation axis. Accordingly, the collective expectation value of the array can be inverted to give an estimate, $\Phi'_{S} = \arcsin{(\langle\sigma_{y}\rangle/C)}$, where $C$ is a scaling factor describing the amplitude of the signal in the absence of injected noise (see supplement). $\Phi'_{S}$ is uniquely defined when the accrued phase lies within $[-\pi/2, \pi/2]$, beyond which the protocol breaks down. 
The estimated noise-induced phase accrued by the data qubits is given by $\Phi'_{D} = (4\beta/3)\Phi'_{S}$, where $4/3$ is the ratio of the sensing times and $\beta = 1.35$ is the ratio of the second-order Zeeman shifts of the clock states (see supplement). With this knowledge, a real-time correction can be applied.

We first probe the case for which the noise is maximally coupled, at $f_{\mathrm{AC}}$ (10.7~mG RMS). Without the spectator protocol, the random phase of the noise leads to complete dephasing of the data qubits. Strikingly, the feed-forward corrects the noise-induced phase in each experimental repetition, resulting in a recovery of the data qubit coherence (Fig.~\ref{fig:Feedback}C). The coherence as a function of the noise amplitude is shown in Fig.~\ref{fig:Feedback}D. In stark contrast to the rapid decay observed in the absence of feed-forward, the spectator protocol robustly preserves coherence for field strengths below 11~mG. Beyond this value, the accrued phases on the spectator qubits can exceed $\pm\pi/2$, where the protocol can no longer unambiguously detect phase errors.

\begin{figure*}[ht!]
\centering
\includegraphics[width = \textwidth]{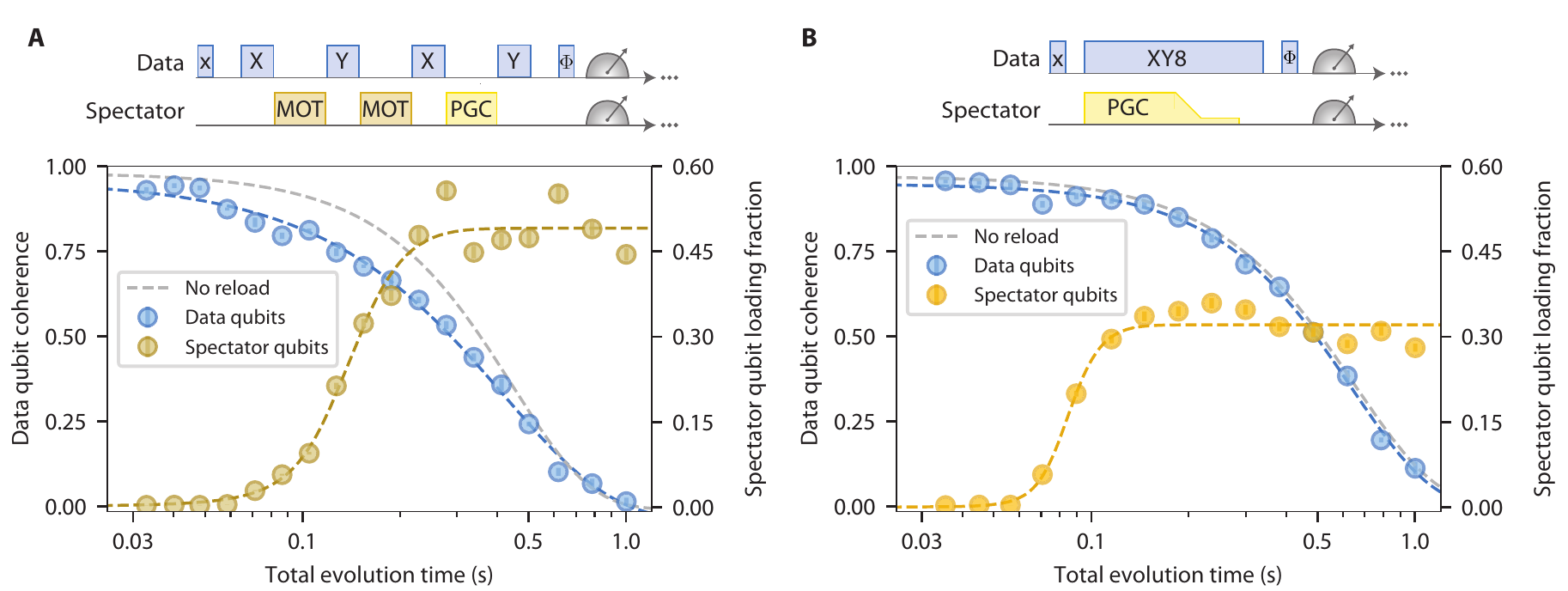}
\caption{\textbf{Reloading of spectator qubits while maintaining data qubit coherence.} (\textbf{A}) Reloading spectators using a pulsed magneto-optical trap (MOT) while decoupling the data qubits. The data qubit coherence time is $T_{2}^{\mathrm{XY4}}$ = 0.42(3)~s with the pulsed MOT and $T_{2}^{\mathrm{XY4}}$~=~0.45(1)~s without it. Spectators are reloaded on a timescale of 150(50)~ms (time required to reach $1-1/e$ of asymptote), saturating at a loading fraction of 0.49. (\textbf{B}) Reloading spectators using polarization-gradient cooling (PGC) during data qubit decoupling. The data qubit coherence time is $T_{2}^{\mathrm{XY8}}$ = 0.64(5)~s with the PGC light, and $T_{2}^{\mathrm{XY8}}$ = 0.65(2)~s without it. Reloading occurs on a faster timescale of 90(30) ms, saturating at a fraction of 0.32. Dashed lines are fits (see supplement).}
\label{fig:Reload}
\end{figure*}

Next, we study the dependence on the noise frequency for an RMS noise strength of 10.7~mG (Fig.~\ref{fig:Feedback}E). For a range of frequencies close to $f_{\mathrm{AC}}$, real-time correction results in an absolute gain in the measured signal, shielding the data qubits from otherwise deleterious decoherence. A pair of small additional features occur near $f_{\mathrm{AC}}$ in the `feed-forward on' spectrum, arising from the finite spectator readout time, which leads to decorrelation between the data and spectator qubits. Reducing the fraction of time used for MCR would suppress these effects. Outside this region, there is a slight reduction in the measured coherence due to imperfect phase estimation. 

For both the amplitude and frequency sweep, the salient features of the data are well described by simple  simulations of the experiment with no free parameters aside from a global amplitude rescaling. These are based on the assumption of monochromatic noise that solely perturbs the frequencies of the qubits (see supplement). At stronger noise strengths, a slight discrepancy occurs, which likely arises from a breakdown of these assumptions. The simulations give insight into the protocol performance. The phase estimation accuracy is limited primarily by uncorrelated dephasing of the spectator qubits (giving $C < 1$), and shot-noise of the spectator readout results, which reduce both the signal-to-noise ratio and the phase resolution (see supplement). This dephasing is likely due to thermal motion in the optical tweezers and tweezer-induced $T_{1}$ processes. Thermal motion can be reduced by additional cooling schemes and $T_{1}$ can be improved by increased detuning of the optical tweezers. Additionally, a larger number of spectator qubits could be used to reduce the shot noise. 
While here we focus on magnetic field noise, the protocol can also mitigate common-mode control errors. For instance, by co-trapping the data and spectator qubits using the same laser system (such as a far-detuned 1064~nm laser), phase errors induced by intensity fluctuations of the trapping laser light could be corrected.

\subsection*{Reloading of spectator qubits}

In these experiments, fluorescence-based detection of the spectators involves selectively removing those in the $\ket{1}_\text{Cs}$ state prior to imaging. Therefore, performing repetitive MCRs will continuously deplete the array. While low-loss readout techniques exist \cite{Fuhrmanek2011,Gibbons2011}, finite losses always remain from both the readout itself and the trapping lifetime. Therefore, continuous operation of atom-based quantum processors will require reload and reset operations which overcome these erasure errors~\cite{Cong22,Wu2022}. Here, we explore two methods for reloading spectators while maintaining coherent data qubits. These build on our standard procedure, where a two-dimensional magneto-optical trap generates a beam of atoms that is laser-cooled into the tweezer array via a three-dimensional magneto-optical trap (MOT).

The first reloading approach uses a stroboscopic MOT that is applied synchronously with an XY4 sequence on the data qubits, to decouple them from the magnetic field gradient (Fig.~\ref{fig:Reload}A). Without the gradient, this decoupling sequence gives $T_{2}^{\mathrm{XY4}}$ = 0.45(1)~s. With it, we find $T_{2}^{\mathrm{XY4}}$ = 0.42(3)~s, but the functional form is modified (see supplement). The spectator array is reloaded on a much shorter timescale of 150(50) ms, defined as the time taken to reach $1-1/e$ of the asymptotic loading fraction. The pulsed MOT saturates at a loading fraction of 0.49, comparable to that achieved with the standard procedure. Residual dephasing from the field gradient can be overcome by using low inductance coils with faster switching times, and by performing decoupling pulses using a Raman laser system, which would enable $\sim$ MHz Rabi frequencies. 

In the second approach, we use polarization-gradient cooling (PGC) to load spectators directly from the atomic beam without a field gradient (Fig.~\ref{fig:Reload}B). This both increases the loading speed and allows an arbitrary choice of decoupling parameters: here we use a single cycle of XY8. In this reloading paradigm, the data qubit coherence time ($T_{2}^{\mathrm{XY8}}$ = 0.64(5) s) is unchanged from the values presented in Fig.~\ref{fig:MCR}, while the spectator qubit array is reloaded on a timescale of 90(30) ms. The fraction of total reloaded spectators is lower than in the previous method, saturating at 0.32. We hypothesize that this is limited by the 2~mm diameter cooling beams. Incorporating larger cooling beams will likely increase the loading fraction for both approaches, and would enable reloading times of a few tens of milliseconds \cite{Steane1992}. Coherence times of $\sim$ seconds can be achieved by using further detuned trapping light and a larger number of decoupling pulses~\cite{Bluvstein22}.

\section*{Discussion}
\vspace{-5pt}

A central challenge for all quantum architectures is to increase system sizes while maintaining low physical error rates. Our demonstration of the use of spectator qubits to measure and correct correlated phase noise is a broadly applicable strategy that can be employed to reduce error rates in quantum computing platforms. Furthermore, spectator protocols could be used in conjunction with standard quantum error correction strategies to protect against correlated errors as well as increase the fidelity of operations beyond the fault-tolerance threshold. An attractive feature of this protocol is that it does not necessitate interactions (two-qubit gates), or individual spectator qubit control, reducing hardware complexity. The use of spectator qubits for noise measurements may provide opportunities in quantum sensing and metrology~\cite{Madjarov2019,Norcia2019, Degen17}, and for improving clock coherence within a single device via differential spectroscopy between the data and spectator qubits~\cite{Kim2021}. While here we focus on global noise, arrays of spectator qubits may also enable the detection of spatially varying noise fields which can be suppressed via local qubit addressing~\cite{Gupta20b}. Careful engineering of the spectator qubits and their control sequences may improve protocol performance. For example, spectator qubits could be encoded in states with enhanced or reduced noise sensitivity to increase the phase resolution or the range of tolerable noise \cite{Song2022}. This can be achieved by using non-zero $m_{F}$ states or by entangling the spectator qubits~\cite{Degen17}. 

The methods demonstrated in this work constitute a set of quantum-control techniques that are essential for atom-array quantum processors, including mid-circuit readout, feed-forward operations, and reloading of auxiliary qubits while maintaining quantum data. Combining these capabilities with programmable intraspecies~\cite{Bluvstein22,Graham2022} and interspecies Rydberg gates will enable auxiliary-qubit-assisted measurements as required for quantum error correction~\cite{Morgado2021,Cong22,Wu2022} and efficient preparation of long-range entangled states~\cite{Verresen2021}. These same capabilities also enable the exploration of complex dynamical quantum behavior under continuous observation, including measurement-induced phase transitions~\cite{Li2018}.



\nocite{Levine2019}
\nocite{Shea2020}
\nocite{Yang2022}
\nocite{Steck01}
\nocite{Steck03}

\vspace{-10pt}
\section*{Acknowledgments}
We thank Harry Levine, Jacob Covey, and Aashish Clerk for fruitful discussions and
critical reading of the manuscript. We acknowledge funding from the Office of Naval Research (N00014-20-1-2510), the Air Force Office of Scientific Research (FA9550-21-1-0209), the NSF QLCI for Hybrid Quantum Architectures and Networks (NSF award 2016136), and the Sloan Foundation. This material is based upon work supported by the U.S. Department of Energy Office of Science National Quantum Information Science Research Centers and was supported by an appointment to the Intelligence Community Postdoctoral Research Fellowship Program at the Pritzker School of Molecular Engineering administered by Oak Ridge Institute for Science and Education through an interagency agreement between the U.S. Department of Energy and the Office of the Director of National Intelligence.

\onecolumngrid
\begin{appendices}

\vspace{-1pt}
\section*{Supplementary Information}

\subsection{Experimental setup}
Our experiment was performed on the dual-species atom array system previously described in Ref. \cite{Singh2022}, which has since been upgraded to provide the functionalities employed for the current results. We use a bichromatic magneto-optical trap to cool $^{87}$Rb and $^{133}$Cs atoms and load them into optical tweezer arrays. Two spatial light modulators (Holoeye PLUTO 2 for Cs, PLUTO 2.1 for Rb) are used to generate the required tweezer arrays at 840 nm and 910 nm for trapping Rb and Cs respectively, with trap spacings of 10 $\mu$m. The optical tweezers are held at $\sim$ 1 mK in depth (with trap frequencies of $\omega_r = 2\pi \times 60$ kHz for Cs and $\omega_r = 2\pi \times 100$ kHz for Rb) when loading atoms during the MOT phase or when reloading during quantum protocols. During quantum circuits, the Rb optical tweezers are ramped down to $\sim$ 140 $\mu$K in depth and the Cs optical tweezers are ramped down to $\sim$ 100 $\mu$K.
Optical pumping is used to prepare qubit states in the hyperfine clock manifolds ($\ket{F = 1, m_F = 0} := \ket{0}_\text{Rb}$ and $\ket{F = 2, m_F = 0} := \ket{1}_\text{Rb}$ for Rb; $\ket{F = 3, m_F = 0} := \ket{0}_\text{Cs}$ and $\ket{F = 4, m_F = 0} := \ket{1}_\text{Cs}$ for Cs). For Cs, this involves on-resonance $\pi$-polarized pumping on the D1 line, and $\pi$-polarized repumping on the D2 line, preparing the atoms in the $\ket{1}_\text{Cs}$ state. For Rb, we use the D2 line for both the $\pi$-pump and the $\pi$-repump, preparing $\ket{1}_\text{Rb}$. For both atomic species we measure optical pumping fidelities of $\sim0.85$ which are predominantly limited by tweezer-induced fictitious magnetic fields, laser polarization purity, the rate of off-resonant scattering out of the $m_{F} = 0$ dark state, and the heating rate during optical pumping. Pumping fidelity can be improved in future work by using Raman-assisted pumping schemes \cite{Levine2019}. Additionally, the addition of Raman laser systems will reduce heating rates during optical pumping and enable $\sim$ MHz Rabi frequencies. 

The qubit states are manipulated by global microwave pulses, at 6.8 GHz for Rb (Stanford Research Systems SG384) and at 9.2 GHz for Cs (Rohde \& Schwarz SGS100A). IQ-modulation of the  microwave sources allows for direct control of the phase, frequency, and amplitude of the hyperfine drive, and enables robust single-qubit manipulations. The microwave tones for the two species are combined (Minicircuits ZFSC-2-10G+), amplified (Minicircuits ZHL6G018G020+), and sent to a single home-built microwave horn. The atoms are confined within an in-vacuum Faraday cage that suppresses stray electric fields but also attenuates the microwave radiation, resulting in microwave Rabi frequencies of 2.53(1)~kHz and 6.48(1)~kHz for Rb and Cs respectively. 
For the implementation of the spectator protocol, we opted to use the Rb atoms as data qubits, as the longer measured decoherence times would be more suitable for general quantum information processing. However, in Section 5 we also show that the read-out process of the Rb atoms does not disturb the Cs atoms, such that the roles could be inverted if desired.
\vspace{-10pt}
\subsection{Real-time processing and feedback} 
State detection is performed by pushing out the atoms in the qubit state $\ket{1}$. The presence of the remaining atoms in the array is detected by collecting 40 ms fluorescence images on an electron-multiplying charged coupled device camera (Andor IXON 888). The camera exposure time is reduced to 15 ms for the mid-circuit readout to enable fast imaging while still maintaining high discrimination fidelity (see Figs.~2C,D of the main text). The image of the spectator qubits is processed in real time, taking less than 8 ms after the exposure period ends for the arbitrary waveform generator (Spectrum Instrumentation M4i.6631-x8) to output the feedback pulse to the IQ-ports of the microwave sources. Significantly reduced fluorescence durations could be achieved by integrating a set of imaging beams separate from the MOT light, for example with a retro-reflected $\pi$-polarized probe beam \cite{Shea2020}. The camera readout time can also be reduced to sub-millisecond times by incorporating software to restrict the pixel readout to specified regions of interest, or by replacing the EMCCD camera with faster imaging technology such as an array of avalanche photodiodes or a qCMOS camera. 
\vspace{-10pt}
\subsection{Quantifying discrimination fidelity of the mid-circuit readout}

To characterize how well the mid-circuit readout discriminates between images with an atom either present (1) or absent (0), we define the discrimination fidelity $\eta$:
\begin{align}
    \eta = 1 - \frac{1}{2} [P(0|1) + P(1|0)]
\end{align}
where $P(i|j)$ is the probability of detecting the state $j$ given that it was actually $i$. modeling the atomic fluorescence data as a bimodal Poisson distribution and performing the histogram fitting techniques previously described in Ref. \cite{Singh2022}, we calculate a photon count threshold and extract $\eta$.  Note that this metric captures the quality of the imaging alone, verifying that the 15 ms duration used here is sufficient to resolve the presence or absence of an atom with a fidelity of $\eta = 0.989(5)$. Small additional errors may be incurred in mapping the hyperfine basis $\{\ket{0}_\text{Cs}, \ket{1}_\text{Cs}\}$ to the $\{$`present', `absent'$\}$ basis via a blow-out pulse, but are independent of the imaging duration. For reference, a typical 40 ms image results in $\eta = 0.996(3)$. 

\vspace{-10pt}
\subsection{Ramsey pulse sequences}
\begin{figure*}[h!]
\centering
\includegraphics[width = 0.5\textwidth]{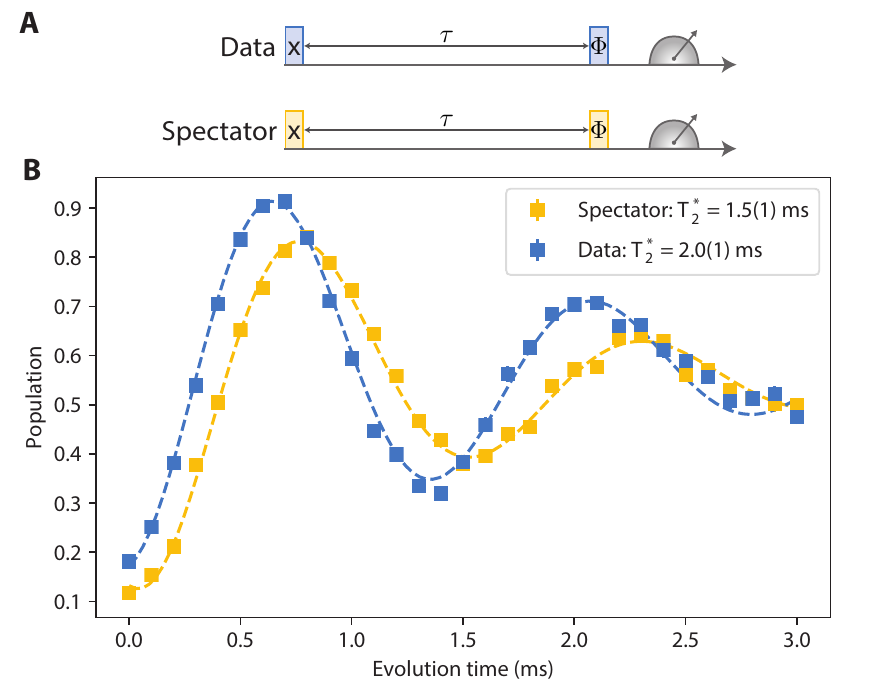}
\caption{\textbf{Ramsey sequence measurements of data and spectator qubits.} (\textbf{A}) Ramsey sequence measurements are performed by applying two $\pi/2$ pulses separated by a period of free evolution time. The phase $\Phi$ of the second $\pi/2$ pulse is varied linearly with the evolution time. (\textbf{B}) $T_2^{*}$ times are extracted from fits of the resulting oscillatory signals (see text). }
\label{fig:Ramsey}
\end{figure*}

The dephasing times of the data and spectator qubits are primarily limited by differential light shifts induced by the optical tweezers. The dephasing times without dynamical decoupling are directly measured using Ramsey pulse sequences (i.e. a sequence consisting of two $\pi/2$ pulses separated by a variable delay time $\tau$) to extract $T_{2}^{*}$. In these sequences, the phase of the second $\pi/2$ pulse is varied linearly with the evolution time to create an artificial detuning. $T_{2}^{*}$ times, shown in Fig.~\ref{fig:Ramsey},  are measured by fitting the resulting oscillatory signals to the following non-exponentially decaying sinusoid: $f(t,A,B,T_{2}^{*},\delta',\phi) = A+B\big[1/(1+1.71 (t/T_{2}^{*})^{2})\big]\cos(\delta't +\phi)$ where $\delta'$ is the artificial detuning \cite{Yang2022}. The coherence times for both the data qubits and the spectator qubits are primarily determined by the choice of wavelength of the optical tweezers used to confine each qubit type. Due to the larger detuning of the optical tweezers that confine the data qubits ($\sim$ 45 nm for the data qubits and $\sim$ 15 nm for the spectator qubits), the coherence time of the data qubits is larger than that of the spectator qubits. Coherence of the spectator qubits can be improved by using further-detuned optical tweezers which would reduce differential light shifts inversely with the square of the detuning, or by cooling the atoms further \cite{Kuhr05}.

\vspace{-10pt}
\subsection{Readout of data qubits while maintaining spectator qubit coherence}
\begin{figure*}[h!]
\centering
\includegraphics[width = 0.45\textwidth]{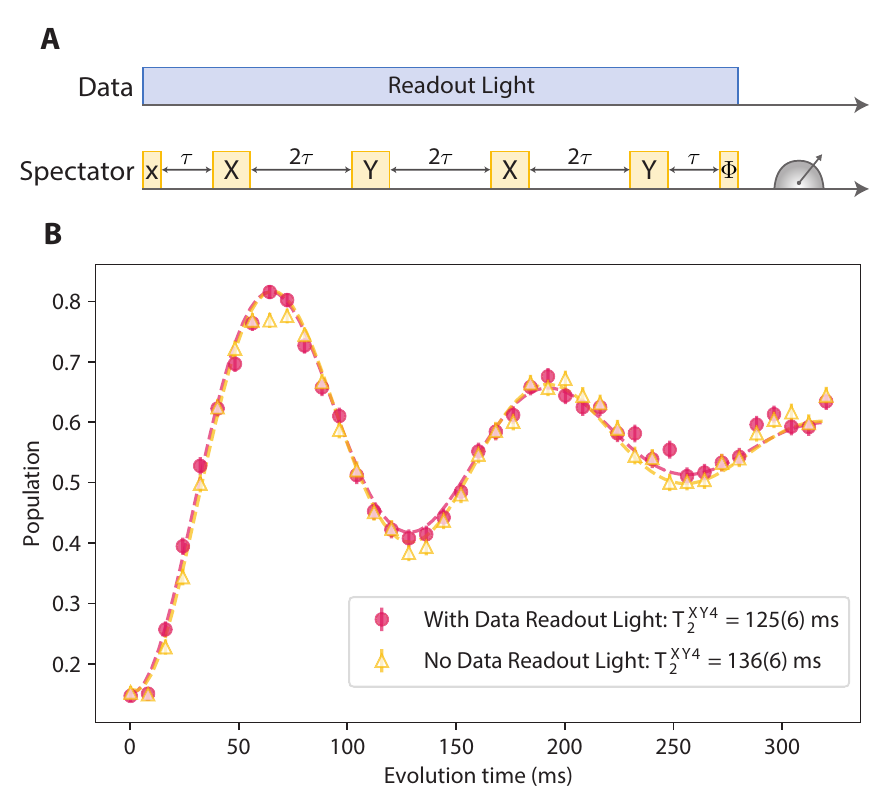}
\caption{\textbf{Mid-circuit readout of data qubits during spectator qubit quantum circuit}. (\textbf{A}) Circuit diagram for measuring the coherence of the spectator qubits during data qubit readout. The readout light of the data qubits is left on for the entire duration of the XY4 decoupling sequence on the spectator qubits. The phase $\Phi$ of the final $\pi$/2 pulse on the spectator qubits is stepped linearly with the total evolution time to create an artificial detuning. (\textbf{B}) Measured population of spectator qubits in $\ket{1}$ as a function of the total evolution time of the decoupling circuit in (A). $T_{2}^{\mathrm{XY4}}$ times are extracted by fitting the oscillatory signals to exponentially-decaying sinusoids. The resulting coherence times remain unchanged with and without the data qubit readout light.    }
\label{fig:RbMCR}
\end{figure*}

The spectator qubits can also be read-out while keeping coherence of the data qubits.
To demonstrate this, we perform an XY4 dynamical decoupling sequence on the spectator qubits (Cs atoms) with and without the readout light of the data qubits (Rb atoms), as shown in Fig.~\ref{fig:RbMCR}. We observe no measurable change of the coherence of the spectator qubits during data qubit readout. The large detuning of the frequency of the data qubit imaging light from the spectator qubit energy levels leads to negligibly small and nearly identical spontaneous scattering rates of $\sim 10^{-7}$ Hz for each spectator qubit level. The calculated $T_1$ time of the spectator qubits if only Raman scattering processes from the data qubit readout light is considered is ~$3\times10^8$ seconds, equating to a spectator qubit bit flip error rate from readout crosstalk of $\sim 5\times 10^{-10}$ during a 40 ms fluorescence image of the data qubits. Because the data qubit readout light is also used for magneto-optical trapping and polarization gradient cooling of the data atoms, the data qubits can in principle be measured or reloaded into the array while maintaining coherence of the spectator qubits. This opens up the possibility that quantum information can be successively swapped between data and spectator qubits between reloading events using Rydberg-based interactions. 

\subsection{Fits of reloading curves}
The number of reloaded spectator qubits in Figs.~4A,B of the main text are fit using the following phenomenological logistic function to extract the spectator qubit reloading parameters: $F(t,A,t_{0},T,n) = A(1-(1+\exp((t-t_{0})/T))^{-n})$. Fit parameters and errors are given in Table~\ref{tab:table1}.
\vspace{-10pt}
\begin{table}[h!]
  \begin{center}
    \begin{tabular}{l|c|c|c|c} 
    $F(t,A,t_{0},T,n)$  &  $A$ & $t_{0}$ (ms) & $T$ (ms) & $n$  \\
      \hline
     MOT Reloading & 0.49(1) & 114(24) & 20(8) & 0.49(39)\\
     PGC Reloading & 0.32(1) & 78(16) & 8(5) & 0.55(74)\\
    \end{tabular}
    \caption{Fit parameters of reloading curves}
    \label{tab:table1}
  \end{center}
\end{table}

\vspace{-20pt}
\subsection{Numerical fits of the coherence}
All coherence measurements in the main text are made by incrementally sweeping the phase of the final $\pi/2$ pulse from 0 to $2\pi$ using IQ modulation of the microwave sources. The amplitude of a cosine fit to the resulting population data as a function of phase angle is used to extract the coherence, namely, $\sqrt{\langle \sigma_{x}\rangle^{2} + \langle \sigma_{y}\rangle^{2}}$. The coherence values are corrected for state preparation and measurement (SPAM) errors using the fitted amplitude of a set of corresponding Rabi oscillations taken prior to each data set. All $T_{2}$ times in the manuscript are measured by fitting the coherence amplitudes as a function of time to the function: $F(t,A,\tau,B,n) = A\exp(-t/\tau)^{n}+B$. The corresponding fit parameters and errors are given in Table~\ref{tab:table2}.

\begin{table}[h!]
  \begin{center}
    \begin{tabular}{l|c|c|c|c} 
    $F(t,A,\tau,B,n)$  &  $A$ & $\tau$ (ms) & $B$ & $n$  \\
      \hline
    Fig. 2B, `No MCR' (Fig. 4B, `No Reload'), XY8 & 0.97(2) & 650(20) & 0.00(2) & 1.71(9)\\
    Fig. 2B, `MCR', XY8 & 0.935(2) & 678(13) & 0.00(1) & 1.82(7) \\
    Fig. 2E, `MCR', XY4 (spectator qubits) & 0.947(5) & 136(7) & 0.00(1) & 1.08(9) \\
    Fig. 4B, `PGC Reload', XY8 & 0.954(7) & 640(50) & -0.01(6) & 1.8(1)\\
    Fig. 4A, `No Reload', XY4 & 0.989(12) & 445(10) & 0.00(1) & 1.79(6)\\
    Fig. 4A, `MOT Reload', XY4 & 1.00(6) & 420(30) & -0.04(5) & 1.3(1)\\
    \end{tabular}
    \caption{Fit parameters of coherence curves.}
    \label{tab:table2}
  \end{center}
\end{table}

\begin{figure*}[h!]
\centering
\includegraphics[width = \textwidth]{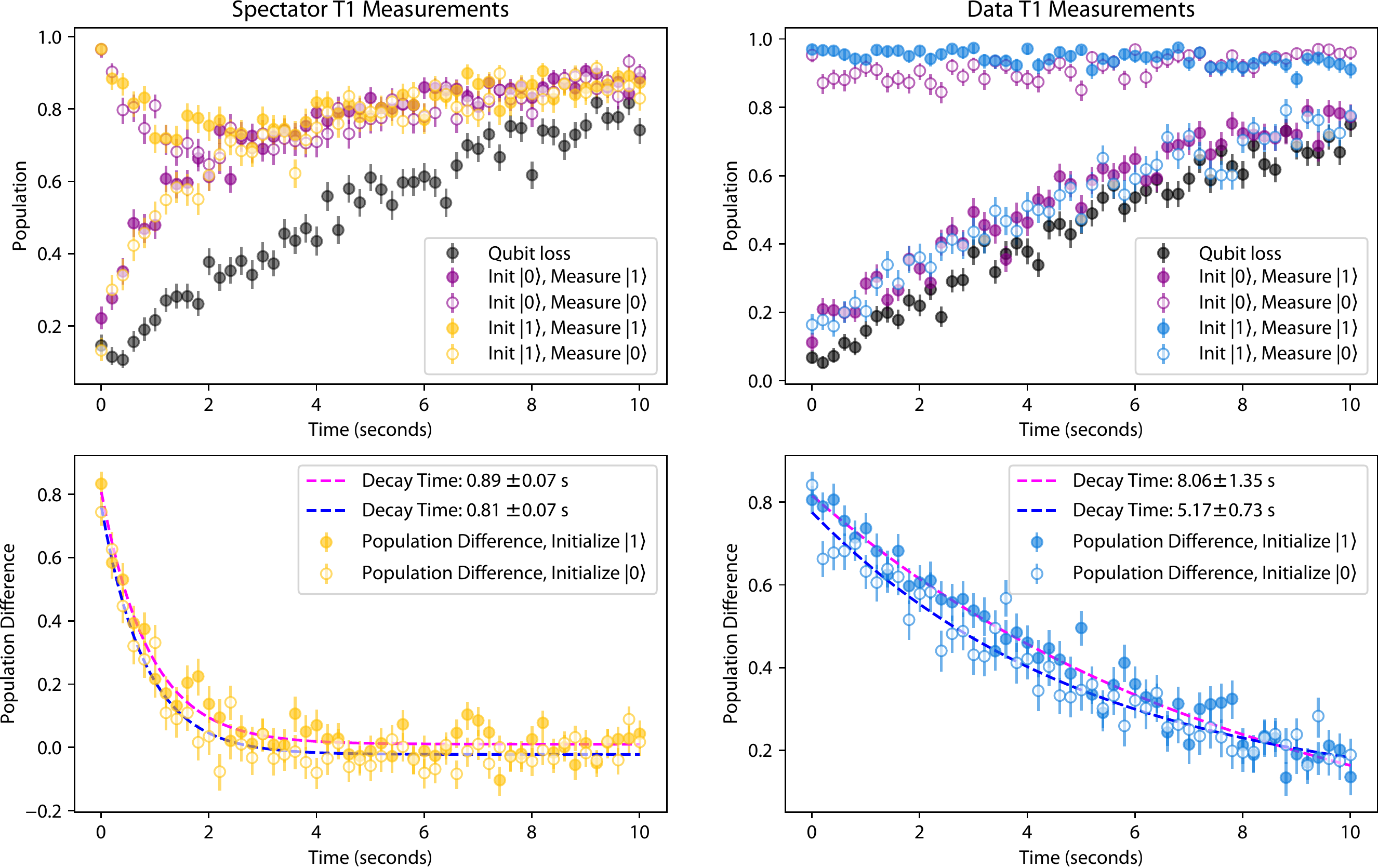}
\caption{\textbf{$\bf{T_{1}}$ measurements of data and spectator qubits}.  $T_{1}$ measurements are performed by initializing the qubits in one of the qubit levels, allowing the qubits to relax for a variable amount of time, and then reading out the qubit population in one of the qubit levels. For each initial state, $T_{1}$ times are extracted by fitting the difference between the final populations measured in each qubit level as a function of time. The average of the fitted decay times results in a $T_1$-type decay time of $0.85(5)$~s for the spectator qubits and $6.6(8)$~s for the data qubits. }
\label{fig:T1Data}
\end{figure*}

\subsection{$T_{1}$ measurements}
Stray laser light, scattering from the optical tweezers, and collisions with the background gas in the vacuum chamber cause population loss from the qubit manifolds on long time-scales. To quantify these effects, we initialize the qubits in either the $\ket{0}$ or $\ket{1}$ level, wait for a varying period of time, and then measure the population either in the $\ket{0}$ state (by selectively removing the population in $\ket{1}$ with a resonant pulse before fluorescence detection, `Measure $\ket{0}$') or in the $\ket{1}$ state (by applying a $\pi$ pulse before removal of the $\ket{1}$ population, `Measure $\ket{1}$'). In Fig.~\ref{fig:T1Data}, we show the data from these measurements. For each initial state, we extract $T_1$-type decay times by fitting the convergence of the population difference between the `Measure $\ket{0}$' and `Measure $\ket{1}$' data. We then average these fitted decay times for each qubit level to extract overall $T_1$-type decay times of $0.85(5)$ seconds for the spectator qubits and $6.6(8)$ seconds for the data qubits. We also directly measure loss of the qubits from the tweezers (grey circles) by performing fluorescence detection of the atoms without selective removal of one of the qubit levels. We measure a background loss rate of $8(1)$ seconds for the spectator qubits and $9(1)$ seconds for the data qubits. The quoted $T_{1}$ times are not corrected for these losses.

\clearpage
\subsection{Modeling the spectator qubit protocol}

In the main text, we investigate the performance of the spectator qubit protocol for magnetic field noise of varying frequencies and strengths. Here we discuss the implementation of the numerical models presented in Figs. 3D,E.

\subsubsection{Overview of the phase estimation protocol}

We first consider the phase estimation routine and describe the choice of feed-forward parameters as used in experiment.

The protocol proceeds as follows. After preparation of the spectator qubits along the x-axis, a noise field induces a phase $\Phi_{S}$ across the decoupling sequence. By measuring the spectator qubits along an orthogonal axis, an estimate for the phase acquired can be made, $\Phi'_{S} = \arcsin({\langle \sigma_{y} \rangle})$, for which a unique value can be assigned while $\Phi_{S} \in [-\pi/2,\pi/2]$. Finally, this can be converted to an estimate for the phase acquired by the data qubits, $\Phi'_{D}$, as discussed below, and fed-forward to the final $\pi/2$ pulse on the data qubits. 

We now consider the experimental implementation. As the spectator qubits are loaded stochastically, the spectator count fluctuates between experimental shots. Therefore, a fluorescence image is taken prior to the start of the quantum circuit to identify the actual number of spectator qubits. The subsequent mid-circuit readout then enables calculation of $\langle \sigma_{y} \rangle$ from the number of those atoms which are found in the bright state, $\langle \sigma_{y} \rangle = 2 (N_{\mathrm{bright}}/N_{\mathrm{initial}}) - 2a$, where $a$ is equal to 0.5 in the absence of SPAM errors.

Ideally, $\Phi'_{S} = \arcsin({\langle \sigma_{y} \rangle})$. In practice, however, the maximal value of $|\langle \sigma_{y} \rangle|$ is reduced by state preparation and measurement errors and uncorrelated decoherence between the spectator qubits. Without compensation for these effects, the accrued phases will be underestimated, leading to sub-optimal feed-forward performance. To account for this, we first characterise the spectator qubit signal in the absence of injected noise. At the end of the 12$\tau$ decoupling sequence, the coherence was found to be $C = 0.46(1)$ (with $a$ as defined above equal to 0.62(1)). The distribution of measurement outcomes from single experimental shots indicates that the data is well approximated by binomial statistics and thus that the decoherence is primarily uncorrelated. This uncorrelated decoherence is likely dominated by variable differential light-shifts arising from the temperature distribution of the atoms \cite{Kuhr05}.

The measured value of $C=0.46(1)$ leads to the phase estimation relations:
\begin{align}
    & \mathrm{if}\  \langle \sigma_{y} \rangle \geq C: \Phi'_{S} = \pi/2 \\
    & \mathrm{elif} \ \langle \sigma_{y} \rangle \leq -C: \Phi'_{S} = -\pi/2 \\
    & \mathrm{else:} \  \Phi'_{S} = \arcsin{(\langle \sigma_{y} \rangle/C)}.
\end{align}

\subsubsection{Calculating the noise-induced phases}

We now consider the noise-induced qubit dynamics, such that the performance of the feed-forward routine can be modeled.

In the experiments, the magnetic field noise is injected along the Z-axis field coils, with strengths of up to 20.5 mG (RMS). The bias field at which the hyperfine qubits are operated is $\{B_{x}, B_{y}, B_{z}\}$ = \{314(1),183(1),357(1)\} mG, giving $|B|$ = 509(1) mG. These values are chosen to be compatible with coherence-preserving reloading of spectator qubits, as they are sufficiently weak to enable MOT formation and PGC in our experiment, but sufficiently strong to maintain the qubit quantization axes. For simplicity, the influence of the noise is treated solely as a time-dependent frequency shift of the qubits. Even for the strongest noise strengths studied here, the maximum tilting of the quantization axis is only a few degrees. Moreover, we assume that the noise is well described by a pure sinusoidal tone at the specified frequency.

The transition frequencies (in Hz) between the $m_F = 0$ clock states of the qubits are modified by second-order Zeeman shifts as:
\begin{equation}\begin{split}
    F_{\mathrm{Rb}} = 6,834,682,611 + 575.15 \times 10^{8} |B|^{2}, \\
    F_{\mathrm{Cs}} = 9,192,631,770 + 427.45 \times 10^{8} |B|^{2},
\end{split}\end{equation}
where $|B|$ is the magnitude of the bias field (in Tesla) \cite{Steck01,Steck03}. The perturbation induced by the noise field, $\delta B_{z} = A \sin(\omega t + \phi)$ is treated as an instantaneous frequency shift:
\begin{equation}\begin{split}
    dF_{\mathrm{i}} = \gamma_{\mathrm{i}} \ ([B_{z}+\delta B_{z}]^{2} - B_{z}^{2}),
\end{split}\end{equation}
where $\gamma_{\mathrm{i}}$ are the second-order Zeeman shifts for each species (Rb:  $575.15 \times 10^{8}$ Hz/T$^{2}$, Cs:  $427.45 \times 10^{8}$ Hz/T$^{2}$). The strength of the applied $\delta B_{z}$ is calculated directly from the control voltage applied to the Z-axis field coil driver using a separately characterized conversion factor of 3.3 mG/mV. 

These instantaneous frequency shifts result in phase accrual by the spectator qubits (Cs atoms, $\Phi_{S}$) and data qubits (Rb atoms, $\Phi_{D}$) across the decoupling sequences. For the spectator qubits, the total evolution time is 12$\tau$ ($[\tau-\pi-\tau]^{6}$), whereas for the data qubits, the total evolution time is 16$\tau$ ($[\tau-\pi-\tau]^{8}$). Each decoupling pulse ($\pi$-pulse) effectively inverts the sign of the frequency shift: at any given time, the effective frequency shift is $(-1)^{n} dF_{\mathrm{i}}$, where $n$ is the number of decoupling pulses which have already been applied. The phases acquired by each qubit type are calculated by integrating the effective instantaneous frequency shifts over these sequences, with $\tau = 6.906$ ms as in the experiment.  As the injected noise is synthesized by a free-running signal generator (Rigol DG812), the phase of the noise signal at the start of the decoupling sequence is unknown and can take any value $\phi_{j} \in$ $[0, 2\pi]$ in each experimental shot.  Therefore, for each noise strength, $A$, and frequency, $\omega$, we numerically evaluate $\Phi_{D}$ and $\Phi_{S}$ for a uniform distribution of $\phi_{j}$ (0.1 degree resolution), from which samples can then be drawn.

\subsubsection{Modeling the phase estimation routine}

Having calculated the noise-induced phases accrued by the two qubit types, we can now estimate the performance of the phase estimation routine using Monte Carlo sampling. For each noise strength, $A$, and frequency, $\omega$, we randomly draw samples $\Phi_{S,j}$ and $\Phi_{D,j}$, corresponding to the calculated phases accrued by the spectator and data qubits for a given initial noise phase, $\phi_{j}$. To account for the effect of shot noise, each sampled $\Phi_{S,j}$ is converted into an associated `atom signal' according to binomial statistics, $f(N, p)$, using the average number of spectator qubits loaded in each experimental shot, $N = 61$, and $p = a + C \sin(\Phi_{S,j})/2$ with $a=0.62$ and $C=0.46$. Running the phase estimation routine on this signal returns an estimated phase $\Phi'_{S,j}$. 

\begin{figure*}[ht!]
\centering
\includegraphics[width = 0.9\textwidth]{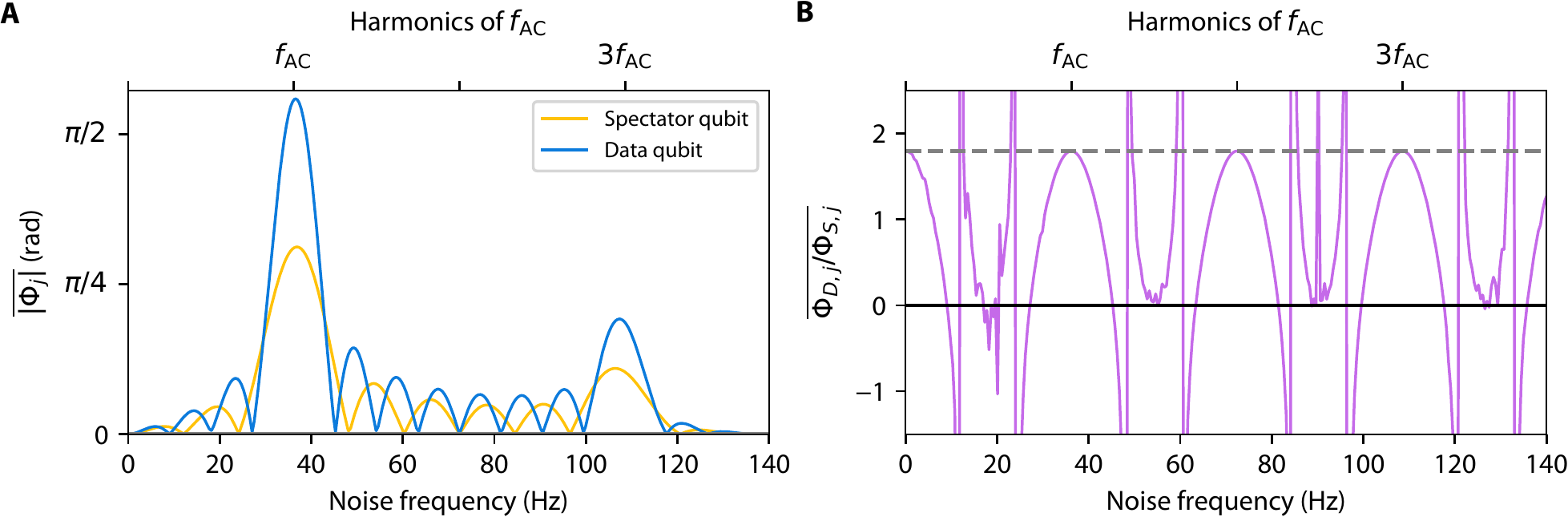}
\caption{\textbf{Phase relationship between spectator and data qubits:} Numerical simulation of the mean phases $\overline{|\Phi_{S,j}|}$, $\overline{|\Phi_{D,j}|}$ (A) induced by 10.7 mG RMS magnetic field noise, and the corresponding phase correlations, $\overline{\Phi_{D,j}/\Phi_{S,j}}$ (B), as a function of frequency. Noise is most strongly coupled into the decoupling sequence at $f_{AC}$ and its odd harmonics. The feed-forward loop is operated with $\Phi'_{D,j} = (4\beta/3)\Phi'_{S,j}$ (grey dashed line in B).}
\label{fig:phaserelation}
\end{figure*}

In the limit of infinitesimally short spectator readout (i.e. identical total evolution times), the phase accumulated by the data qubits would be related to that of the spectators as: $\Phi'_{D,j} = \beta\Phi'_{S,j}$, where $\beta = 1.35$ is the ratio of the second-order Zeeman shifts. In that limit, the spectator protocol is agnostic to the spectrum of the noise. Here, the finite readout time must be taken into account, giving $\Phi'_{D,j} = (4\beta/3)\Phi'_{S,j}$, where the factor $4/3$ arises from the difference in total evolution times. The finite time associated with readout causes a breakdown of the phase relationship for certain frequencies, as exemplified in Fig.~\ref{fig:phaserelation}. However, this relationship does hold for `worst-case' noise which couples maximally into the decoupling sequence, at $f_{AC}= 1/(4\tau) = 36.2$ Hz and its odd harmonics (Fig.~\ref{fig:phaserelation}). We thus use this phase relationship for the construction of the feed-forward loop. 

From the sampled accrued phases, $\Phi_{S,j}$, and corresponding estimates $\Phi'_{S,j}$, the performance of the protocol can be evaluated. This performance is captured by the residual coherence of the data qubits at the end of the quantum circuit. In the experiment, this is measured by varying the phase of the final $\pi/2$ analysis pulse, and fitting the resulting amplitude. That is, when the feed-forward is turned off, each pulse is applied with a phase $\phi_{k}$ from a set of predetermined sweep values, and when the feed-forward is turned on, the pulses are applied with phases ($\phi_{k} + \Phi'_{D,k}$), where $\Phi'_{D,k}$ is the estimated phase for that specific instance of the experiment. Each presented data point in Figs. 3D,E results from the fitted amplitude from a sweep of 10 $\phi_{k}$ values (akin to Fig. 3C), with 80 repetitions per $\phi_{k}$. We evaluate this fit at $\phi_{k} = 3\pi/2$, i.e. $\langle \sigma_{x} \rangle$, in order to capture cases in which the qubit state has not only decohered but is even inverted.

\begin{figure*}[ht!]
\centering
\includegraphics[width = 0.85\textwidth]{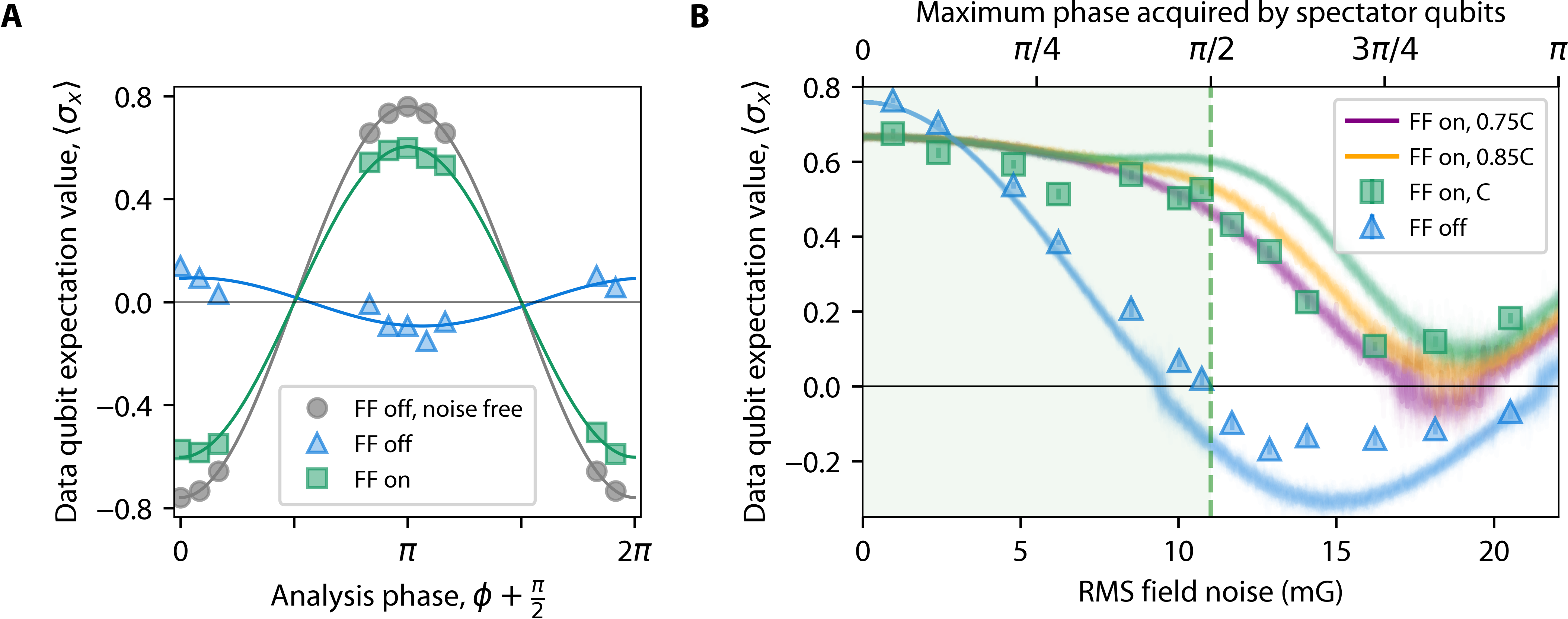}
\caption{\textbf{Simulated feed-forward protocol performance:} A) Numerically generated data qubit expectation values, using the same sweep points as were used for the experimental data presented in Figs. 3D,E of the main text. Here we show examples without feed-forward, both in the absence of injected noise (`FF off, noise free'), and with 10.7 mG RMS noise at $f_{AC}$ (`FF off'). For the latter case we also show the simulated performance when the feed-forward is turned on (`FF on'). The fitted amplitudes are used to extract $\langle \sigma_{x} \rangle$ for each noise strength and frequency. Figs. 3D,E of the main text present the results of 100 such Monte Carlo instances, alongside the experimental data.  B) Additional numerical simulations of the feed-forward protocol performance as presented in Fig. 3D. Alongside the simulated protocol performance considering spectator qubit data generated using $C=0.46$ (as measured in experiment in the absence of noise, and for which the feed-forward parameters are optimized), here we show the performance obtained for $0.75C$ and $0.85C$, in which case the feed-forward would under-correct the accumulated phases.}
\label{fig:simulation}
\end{figure*}

To account for the finite sampling statistics and fitting procedure, each simulation curve presented in Figs. 3D,E is generated following an analogous process. For each noise strength and frequency, we generate 80 samples of \{$\Phi_{S,j}$, $\Phi'_{S,j}$, $\Phi_{D,j}$,  $\Phi'_{D,j}$\} for each of the 10 $\phi_{k}$. When the feed-forward is turned off, each sample results in a modulation of the data qubit expectation value by a factor $f_{\mathrm{off,j}} = \cos(\Phi_{D,j})$. Likewise, when the feed-forward is turned on, it is modulated by a factor $f_{\mathrm{on,j}} = \cos(\Phi_{D,j}-\Phi'_{D,j})$. For each $\phi_{k}$, we thus get overall multiplication factors, $\{f_{\mathrm{off,k}},f_{\mathrm{on,k}}\}$ from the means of the 80 $\{f_{\mathrm{off,j}}, f_{\mathrm{on,j}}\}$. Finally, those factors are applied to an independently measured noise-free characterization curve, $p = D \cos(\phi_{k})$, and the resulting amplitude is fitted. For the simulations presented in Fig. 3E of the main text, both curves are rescaled by an additional common factor of 0.93, obtained from a least-square fit to the `feed-forward off' data, which we attribute to day-to-day fluctuations of the optical pumping fidelity and Rabi frequencies. Examples of generated noise-free, `feed-forward off' and `feed-forward on' datasets are shown in Fig.~\ref{fig:simulation}A.

\subsubsection{Performance of the feed-forward protocol}

In this section we discuss in further detail the performance of the feed-forward protocol, and avenues for improvement.

First, we consider the dependence on the amplitude of the noise. The key behaviour is that, when the feed-forward is turned off, any resonant noise at $f_{AC}$ causes decoherence of the data qubits. As a function of noise strength, the coherence exhibits a decaying oscillatory behaviour. These oscillations arise from the extrema of the distribution of $\Phi_{S}$ induced by a given noise strength, for which the associated modulation factors $f_{\mathrm{off}}$ can push the expectation value above or below 0. 

 Conversely, when the feed-forward is turned on, the system first exhibits robustness to noise of up to 11 mG, the quintessential behaviour of the protocol. Beyond this point, the extrema of the $\Phi_{S}$ distribution exceed $\pm\pi/2$, and the protocol begins to fail. The experimental implementation is seen to break down slightly earlier than predicted by the no-free-parameter model. Such an effect can be phenomenologically reproduced in the model by under-correcting for the $\Phi_{D}$ (Fig.~\ref{fig:simulation}B).  The slight disagreement which also occurs in this region for the `feed-forward off' data indicates that the simulated $\Phi_{S}$ do not perfectly match experiment. Such a discrepancy may arise from some non-linearity in the synthesis and injection of the noise, or from more complex qubit dynamics than simple frequency shifts. Note that finite pulse durations and pulse errors are also not captured by the model.


Next, we turn to the features of the frequency sweep. As expected, the feed-forward protocol largely suppresses the effects of noise which would otherwise enter the decoupling sequence, which occur at  $f_{AC}$ and its odd-harmonics \cite{Degen17}. In the present implementation, however, the protocol also introduces some additional sensitivity to noise at the edge of these features, resulting in two small `dips'. These arise from the finite time associated with readout of the spectator qubits. As discussed above, and shown in Fig.~\ref{fig:phaserelation}, the phase relationship $\Phi'_{D,j} = (4\beta/3)\Phi'_{S,j}$ is not valid for all frequencies. In such cases, the feed-forward loop will either under- or over-correct phases acquired by the data qubits. This effect can be mitigated by reducing the fraction of the time associated with spectator readout. Regardless, the protocol successfully suppresses noise which would otherwise completely decohere the data qubits.

Alongside these features, for both the amplitude and frequency sweeps, there is a modest reduction in the data qubit coherence across the entire range, arising from imperfect phase estimation. In general, the quality of the phase estimation routine is dependent on both the number of available spectator qubits, and the amount of information which can be extracted from each spectator qubit. There are thus two ways to generally enhance the performance. First, one can simply increase the number of Cs atoms. Second, the value of $C$, the mutual coherence of the spectator qubits at the end of the sensing sequence, can be improved. 

At present, $C$ is limited by two main factors. The quality of optical pumping limits the state preparation fidelity for the spectator qubits to $\sim$90\%. This could be improved by using Raman-assisted state preparation schemes \cite{Levine2019}. Moreover, the spectator qubits undergo significant uncorrelated dephasing in the $\sim$110 ms sensing time. Increasing the detuning of the optical tweezers from the Cs D1 and D2 lines would suppress both unwanted $T_{1}$ processes and dephasing from differential light-shifts, as observed for the rubidium qubits. 

For the experimental parameters used in this work  ($N=61$ spectator qubits and $C=0.46$), the numerical simulations predict that imperfect phase estimation causes a reduction of the data qubit $\langle \sigma_{x} \rangle$ by $f_{\mathrm{on}}=0.880$ in the absence of any injected noise. With $C=1$, this factor would immediately be improved to $f_{\mathrm{on}}=0.974$; that is, the majority of the error comes from limited spectator contrast as opposed to the shot noise limit from the number of spectator qubits. Increasing to $N=165$ would result in a factor of $f_{\mathrm{on}}=0.956$ for $C=0.46$, or $f_{\mathrm{on}}=0.990$ for $C=1$, i.e. a 0.5\% error in the state fidelity. 

Note that here, the data qubits are slightly more sensitive to the noise field than the spectator qubits. As discussed in the outlook of the main text, further improvements in performance may be achieved for specific applications by careful tailoring of the spectator qubits. For example, encoding in states with increased sensitivity to the noise field (e.g. non-zero $m_{F}$ states), or using entangled spectator states \cite{Song2022,Degen17}, can significantly mitigate errors from imperfect phase estimation, at the cost of a reduced range of correctable noise. This would likely be desirable when concatenating spectator protocols with quantum error correction.

\end{appendices}

\clearpage

\end{document}